# MinSurf: resolving the atomic-scale stability landscape of mineral surfaces

*Fengzijun Pan[a,b], Zhoulin Liu[c,d], Pingyang Zhang[d], Jiaqiu Xu[e], Zepeng Fan[a,b,*],*

*Dawei Wang[a,b], Jianzhong Pei[f],*

[a] School of Transportation Science and Engineering, Harbin Institute of Technology, Harbin, China

[b] Chongqing Research Institute of HIT, Chongqing, China

[c] School of Science, Harbin Institute of Technology, Shenzhen, China

[d] National Engineering Laboratory for Reducing Emissions from Coal Combustion, Shandong University, Jinan, China

[e] Department of Civil and Environmental Engineering, The Hong Kong Polytechnic University, Hong Kong, China

[f] School of Highway, Chang'an University, Xi'an, China

*Email: zepeng.fan@hit.edu.cn

**Abstract:** Mineral surfaces govern interfacial reactivity in carbon mineralization, geo-energy storage, contaminant immobilization, heterogeneous catalysis and electrochemical interface engineering. Yet atomistic simulations often rely on commonly used facets or facet-level stability criteria, while distinct atomic terminations of the same crystallographic orientation are rarely resolved systematically because experimental characterization and density functional theory (DFT) calculations remain costly across large surface spaces. Here we present MinSurf, a high-throughput framework that resolves mineral surface selection as a surface-energy and morphology problem. MinSurf integrates surface enumeration, DFT labelling, machine-learning interatomic potentials and Wulff construction to predict stable terminations, surface-energy landscapes and equilibrium crystal morphologies. Applied to ten representative minerals, MinSurfSet comprises 764 surface slabs, with 90 corresponding oriented unit cells constructed as bulk references for surface-energy evaluation. The resulting MinNEP model predicts DFT surface energies with a mean absolute error of 0.0119 eV $Å^{-2}$ and achieves an overall acceleration of $1.14 \times 10^4$ relative to DFT. MinNEP preserves the DFT-derived morphology-determining surface-energy hierarchy and reproduces the dominant Wulff-exposed facets, while X-ray diffraction provides an independent crystallographic consistency check for the α-quartz benchmark. By linking atomic terminations, surface energies and equilibrium morphologies, MinSurf provides reproducible and physically representative surface models for high-throughput simulations of mineral interfaces across energy, environmental and advanced inorganic materials.

## Introduction

Mineral surfaces mediate matter exchange, charge redistribution and interfacial bonding across natural, engineered and planetary materials[1–3]. At exposed crystal facets, local atomic coordination governs how water, ions, molecules and binding phases interact with mineral frameworks, thereby regulating hydration, dissolution, adsorption, ion exchange, nucleation, carbon mineralization and contaminant immobilization[4–6]. These surface-mediated processes underpin construction materials, environmental interfaces, energy technologies, catalytic systems and planetary regoliths and advanced inorganic materials[7,8]. Understanding mineral surfaces at the atomic scale is therefore essential for linking crystal structure to interfacial reactivity, morphology evolution and material performance[6,9].

Atomistic simulation provides a route to establish this link at the scale where mineral surfaces are defined[10,11]. By resolving local coordination, surface relaxation, charge redistribution and molecular interactions near exposed facets, it can reveal structure–property relationships that are difficult to access experimentally[12]. This capability is becoming increasingly important for materials design, where large chemical and structural spaces must be explored to identify stable, reactive or functional interfaces. High-throughput atomistic modelling therefore offers a path from case-by-case surface studies towards systematic prediction of mineral interface behaviour[13]. Its predictive value, however, depends on whether the surface models used in these simulations represent physically relevant exposed surfaces.

This requirement creates a less visible but consequential uncertainty: the choice of the surface slab. In many atomistic studies, surface models are inherited from previous work, inferred from common cleavage planes, selected using facet-level stability arguments or chosen for computational convenience[14]. Such choices may appear technical, but they can determine predicted adsorption strength, dissolution pathway, hydration mechanism, charge distribution and interfacial adhesion[15]. The central difficulty is that a crystallographic facet does not uniquely define an atomic surface[8]. A Miller index specifies the orientation of a plane, but not the termination exposed after cleavage. The same facet can generate multiple non-equivalent atomic surfaces with different stoichiometry, polarity, coordination environment, relaxation behaviour and surface energy, especially in minerals with mixed ionic–covalent bonding, low symmetry, large unit cells and heterogeneous cation sublattices[15–17].

A rigorous mineral surface model should therefore be placed within the broader surface landscape of a crystal rather than defined by a single cleavage plane[6,15]. This landscape includes possible crystallographic orientations, atomic terminations, relaxation responses, relative stabilities and morphological exposures. Surface energy provides a thermodynamic coordinate for comparing candidate surfaces, while Wulff construction translates this energetic landscape into the facets expected to remain exposed at equilibrium[16,18]. Resolving this landscape remains difficult at scale. Experimental techniques such as X-ray diffraction (XRD) and high-resolution transmission electron microscopy (HRTEM) provide essential structural information, but they are not readily scalable for large numbers of candidate facets and terminations[19,20]. Density functional theory (DFT) provides a first-principles route to surface relaxation and energy evaluation, yet its cost is intrinsically high because each structural optimization step requires repeated self-consistent solution of the electronic structure[21]. Machine-learning interatomic potentials (MLIPs) offer a route to bridge this gap, but mineral surfaces require models that can describe bulk-like interiors, under-coordinated surface atoms, distorted coordination polyhedra, polar environments and small surface-energy differences across competing terminations[22].

Here we present MinSurf, an extensible workflow for resolving stable and morphologically exposed mineral surfaces using machine-learning interatomic potentials. MinSurf integrates surface enumeration, density functional theory labelling, relaxation with MinNEP—a charge-aware neuroevolution potential (qNEP) tailored to mineral surfaces—and Wulff construction to connect atomic slab models with surface-energy landscapes and equilibrium crystal morphologies. MinSurfSet is constructed as a slab-based first-principles dataset in which surface slabs are paired with corresponding oriented unit cells used as bulk references for surface-energy evaluation. Trained on this dataset, MinNEP captures mineral surface energetics through environment-dependent effective charges and first-principles supervision of energies and atomic forces. By moving beyond convention-based or facet-level surface selection, MinSurf provides a standardized and expandable route for constructing physically grounded surface models for high-throughput simulations of mineral interfaces.

## Results

### MinSurf workflow for resolving stable mineral surfaces

Predictive mineral simulations require surface models that are selected according to stability rather than convention. However, mineral slabs are often inherited from previous studies, inferred from common cleavage planes or chosen for computational convenience, even though such choices may not uniquely define the physical surface being simulated. The same Miller index can correspond to different atomic terminations, stoichiometries or polar compensation schemes, leading to distinct surface energies and relaxation behaviours[15,23]. Because adsorption, dissolution, hydration and interfacial bonding are inherently surface-specific, this ambiguity can propagate directly into the predicted mechanisms and interfacial properties[9,15].

MinSurf addresses this issue by treating surface construction as a reproducible surface-resolution task. The workflow begins from a mineral crystal structure and generates OUCs along selected Miller indices (**Fig. 1a**). For each crystallographic orientation, candidate terminations are enumerated rather than assuming a single surface for each facet. This separates the orientation of a plane from the atomic termination actually exposed to the environment, allowing competing surface models to be compared using a common thermodynamic criterion.

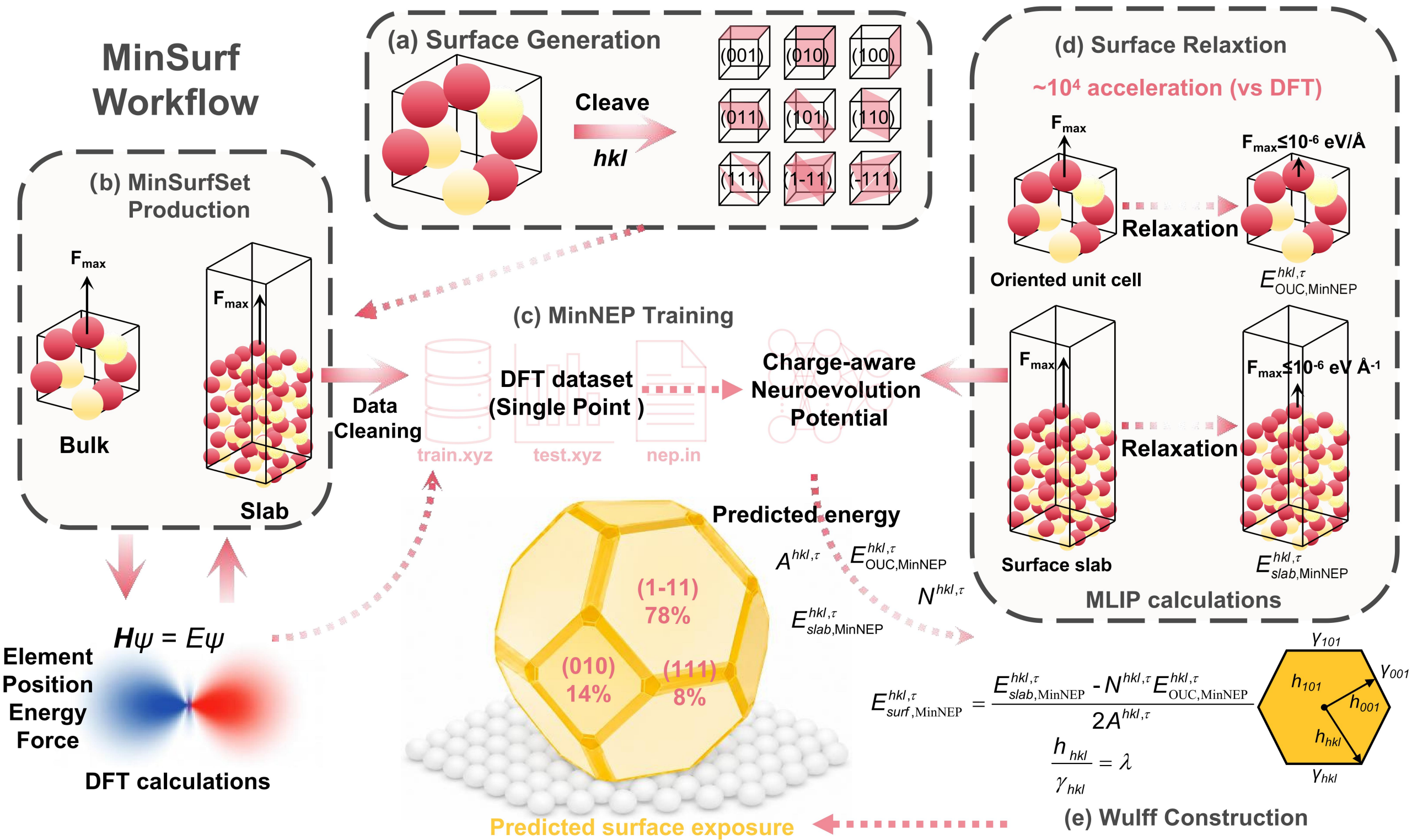

**Fig. 1 | MinSurf workflow for resolving stable mineral surfaces**. **a,** Generation of candidate surfaces by cleaving OUCs along selected Miller indices. **b,** Construction of MinSurfSet from DFT-labelled slab configurations with corresponding OUCs references. **c,** Training of the MinNEP model using the curated MinSurfSet. **d,** Rapid relaxation of OUCs references and surface slabs using MinNEP, enabling an overall speedup of $1.14 \times 10^4$ relative to the DFT reference workflow. e, Wulff construction based on predicted surface energies to determine equilibrium morphology and surface exposure.

The generated OUCs and slab configurations are labelled using DFT to provide reference energies and atomic forces for model development (**Fig. 1b**). These data are assembled into MinSurfSet and used to train MinNEP, a charge-aware MLIP designed for mineral surface environments (**Fig. 1c**). Once trained, MinNEP enables rapid relaxation of OUCs and surface slabs and provides the energies required to rank competing facets and terminations at substantially reduced computational cost relative to DFT (**Fig. 1d**).

The final step converts surface energies into surface exposure. Surface energies calculated from relaxed bulk-reference and slab structures are used for Wulff construction, which maps orientation-dependent stability onto equilibrium morphology and surface area fraction (**Fig. 1e**). MinSurf therefore provides a standardized route from crystal structure to thermodynamically stable and morphologically exposed mineral surfaces. Because newly added minerals, facets and terminations can be incorporated using the same generation and labelling logic, the workflow is inherently extensible and can support a growing surface-energy landscape for mineral modelling.

**Construction of MinSurfSet and training of MinNEP**

To provide first-principles supervision for mineral surface energetics, we constructed MinSurfSet from ten representative mineral crystals spanning calcium silicates, calcium aluminates, aluminosilicates, feldspars and quartz (**Supplementary Table S1** and **Supplementary Fig. S1**)[24–26]. These minerals cover representative phases relevant to broad surface-controlled applications, including low-carbon cementitious materials, geochemical and environmental mineral reactions, and advanced inorganic functional materials, while retaining broader chemical and crystallographic diversity[27–29]. For each crystal, OUCs and surface slabs were generated from selected Miller indices and non-equivalent terminations, so that both bulk-like coordination environments and under-coordinated surface configurations were included (**Fig. 2a**)[30]. The procedures for OUC construction, slab generation and termination enumeration are described in the **Methods**.

MinSurfSet contains 90 OUCs and 764 surface structures. The number of generated surfaces varies substantially among minerals, reflecting differences in crystal symmetry, primitive-cell size and termination diversity. This structure-dependent distribution highlights the need for systematic enumeration: low-symmetry or chemically complex minerals can generate many possible slabs, whereas highly symmetric phases may produce only a small number of distinct surfaces.

As detailed in the **Methods**, DFT calculations were performed using parameter settings validated by systematic convergence tests to provide reliable reference energies and atomic forces. In total, the labelling workflow produced 75,906 DFT-labelled configurations, including surface slabs and their corresponding OUC references, of which 75,305 converged calculations were retained as valid reference labels after data cleaning (**Supplementary Table S2**). This reference labelling strategy is essential for surface-energy prediction, because surface energy is evaluated from the energetic difference between a relaxed slab and its corresponding OUC reference.

MinNEP adopts the qNEP framework, in which local atomic energies are combined with environment-dependent effective charges and a reciprocal-space electrostatic term, allowing long-range electrostatic contributions to be retained while preserving the efficiency of NEP (**Supplementary Section S1**)[23,31]. The final model was trained for Ca, Si, O, Na, Al, Fe and K using 7 Å radial and angular cutoffs, a 40-neuron hidden layer and a descriptor basis size of 12. These settings were selected from hyperparameter tests comparing surface-energy accuracy, energy and force errors, model size and training cost (**Supplementary Section S6**).

The trained model was then evaluated against DFT reference energies and forces and compared with a conventional NEP baseline.

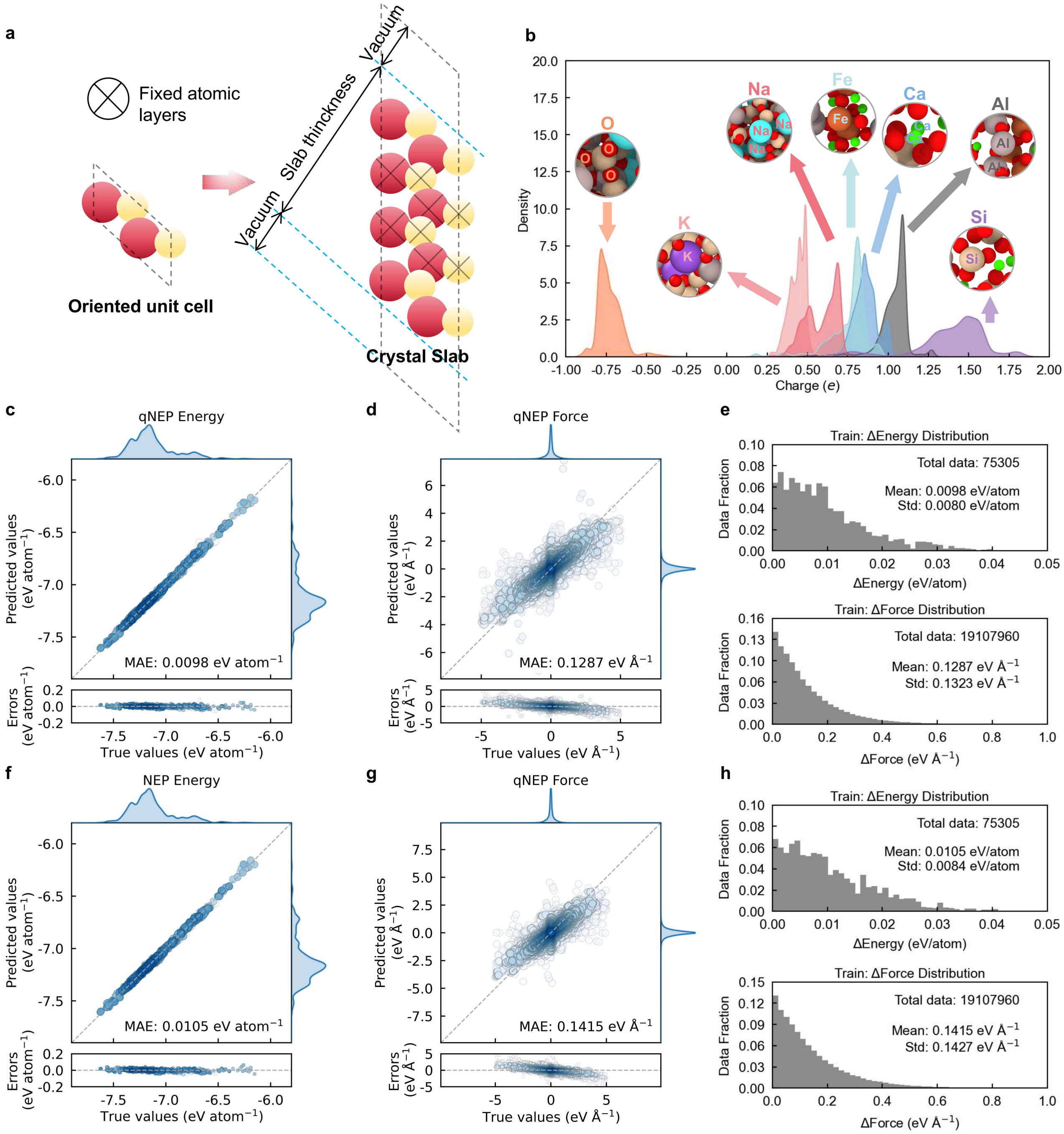

**Fig. 2 | Slab construction and model-level evaluation of MinNEP. a,** Generation of surface slabs from OUCs with vacuum layers and fixed atomic layers. **b,** Learned effective charge distributions of different elements in the charge-aware qNEP model. **c,d,** Parity plots of qNEP-predicted energies and forces against DFT reference values. **e,** Error distributions of NEP for energy and force predictions. **f,g,** Parity plots of conventional NEP-predicted energies and forces. **h,** Error distributions of conventional NEP for energy and force predictions.

## Evaluation of the MinNEP model

Before assessing surface-energy prediction, we first evaluated whether MinNEP provides a reliable potential energy surface for mineral slab relaxation. This model-level evaluation is important because surface energy depends not only on final total energies, but also on the forces that determine how surface atoms relax after cleavage[8]. The DFT-labelled configurations used for model development contain surface-specific environments, including under-coordinated atoms, distorted coordination polyhedra and relaxation-induced local structural changes.

The learned effective charges show clear element-dependent distributions across O, Na, K, Fe, Ca, Al and Si environments (**Fig. 2b**). Oxygen atoms are mainly assigned negative effective charges, whereas network-forming and charge-balancing cations occupy positive charge ranges. These learned charges should not be interpreted as fixed oxidation states. Instead, they are environment-dependent latent variables optimized to reproduce DFT energies and forces within the qNEP framework. Their chemically separated distributions indicate that MinNEP captures distinct local bonding environments while retaining the flexibility to adapt charges during structural relaxation.

MinNEP accurately reproduces DFT energies and forces for the labelled configurations. The energy parity plot shows close agreement with DFT reference values, with a mean absolute error of 0.0098 eV atom$^{-1}$ (**Fig. 2c**). Force prediction is more demanding because surface cleavage and relaxation generate local environments that deviate from ideal bulk coordination. Nevertheless, MinNEP achieves a force mean absolute error of 0.1287 eV Å$^{-1}$, and the corresponding error distribution remains concentrated near zero (**Fig. 2d,e**). This force accuracy supports the use of MinNEP for relaxing candidate mineral slabs before surface-energy evaluation.

We further compared the charge-aware qNEP with a conventional NEP trained using the same data. The conventional NEP captures the overall energy and force trends, but gives slightly larger errors, with an energy mean absolute error of 0.0105 eV atom$^{-1}$ and a force mean absolute error of 0.1444 eV Å$^{-1}$ (**Fig. 2f–h**). The improvement is more evident for forces than for energies, suggesting that the charge-aware formulation is particularly useful for describing the local gradients of the potential energy surface. This behaviour is consistent with the partially ionic character of the mineral systems considered here, where polar environments and coordination changes can introduce electrostatic contributions beyond a purely local description. These results establish the model-level basis for the subsequent evaluation of surface-energy prediction and morphology-resolved surface exposure.

**Accuracy and efficiency of MinNEP for surface-energy prediction**

Surface-energy prediction is the central test of whether the MinNEP framework can resolve stable mineral surfaces, because it couples slab relaxation, OUC reference consistency and the relative energetic ordering of competing terminations. Unlike total energy or force prediction, surface energy is evaluated from the energy difference between a relaxed slab and its corresponding OUC reference. Even small errors in relaxation or energy evaluation can therefore affect the surface-energy hierarchy, which ultimately determines whether a facet persists or disappears in Wulff construction. Two MinNEP variants trained on MinSurfSet, the qNEP version and the conventional NEP version, were therefore evaluated against DFT-calculated surface energies and compared with the previously reported NEP89 model (**Fig. 3a–c**)[32].

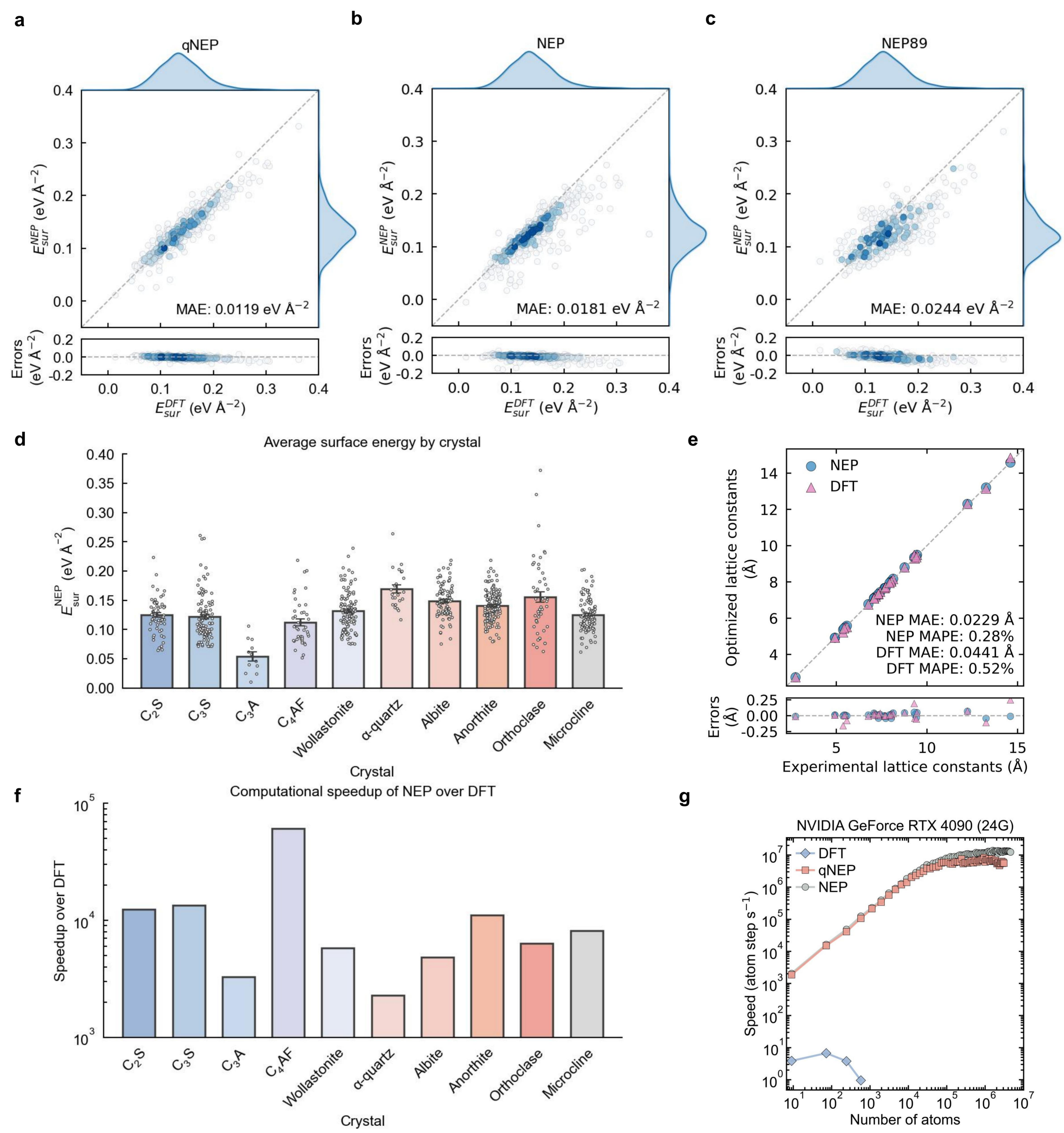

**Fig. 3 | Accuracy and efficiency of MinNEP for mineral surface prediction. a–c,** Parity plots comparing DFT surface energies with predictions from two MinSurfSet trained MinNEP variants, namely the charge aware qNEP version and the conventional NEP version, together with the previously reported NEP89 model. **d,** Mineral resolved surface energy distributions predicted by the charge aware qNEP version, with bars indicating

mean values and points representing individual slab terminations. **e,** Comparison of qNEP and DFT optimized lattice constants with experimental values. **f,** Computational speedup of the charge aware qNEP version relative to the DFT reference workflow for surface energy calculations of different minerals. **g,** Scaling of computational speed with system size on an NVIDIA GeForce RTX 4090 GPU.

The qNEP version gives the closest agreement with DFT surface energies, with a mean absolute error of 0.0119 eV $Å^{-2}$ (**Fig. 3a**). The conventional NEP version gives a larger error of 0.0181 eV $Å^{-2}$, while NEP89 gives an error of 0.0244 eV $Å^{-2}$ (**Fig. 3b,c**). Because NEP89 is a broad 89 element model designed for general atomistic environments, this comparison indicates that broad elemental coverage alone is not sufficient for resolving mineral surface energetics. The improved performance of the MinSurfSet-trained qNEP version shows that surface-energy prediction benefits from slab data targeted to mineral surface environments and from an explicit treatment of environment-dependent charges. This is consistent with the nature of cleaved mineral surfaces, where polarity, under-coordination and local charge redistribution can affect the relative stability of competing terminations.

The mineral-resolved surface-energy distributions further show that surface stability is controlled by both mineral chemistry and termination structure (**Fig. 3d**). Different minerals exhibit distinct average surface-energy levels, reflecting differences in crystal framework, bonding character and exposed cation environments. At the same time, the spread of individual slab terminations within each mineral shows that termination chemistry can be as important as crystallographic orientation. Surfaces generated from the same parent crystal may expose different cation–oxygen polyhedra, coordination defects or polar arrangements, leading to different energetic penalties. This result explains why explicit termination enumeration is necessary: a conventionally selected slab may not correspond to the most stable or morphologically relevant surface.

The qNEP version also preserves the structural consistency required for reliable surface-energy evaluation. Its optimized lattice constants agree closely with experimental values, with a mean absolute error of 0.0229 Å and a mean absolute percentage error of 0.28% (**Fig. 3e**)[25]. In the same comparison, DFT gives a mean absolute error of 0.0441 Å and a mean absolute percentage error of 0.52%. This result should be interpreted as a structural consistency check rather than as a general claim that qNEP is intrinsically more accurate than DFT. The agreement indicates that the model maintains the equilibrium geometry of mineral frameworks during relaxation, which can be attributed to its ability to reproduce the local force balance around cation–oxygen coordination polyhedra while retaining charge-dependent electrostatic constraints in partially ionic lattices. This supports its use for optimizing both OUC references and surface slabs.

The efficiency of the qNEP version changes the practical scale of surface-energy evaluation. It achieves an overall speedup of $1.14 \times 10^4$ relative to the DFT reference workflow (**Fig. 3f**), because energies and forces are evaluated directly from learned atomic descriptors and environment-dependent charges rather than through repeated electronic self-consistency iterations. The magnitude of acceleration varies among minerals because the computational burden of DFT is strongly affected by crystal structure, cell size, elemental composition, electronic complexity and the number and complexity of slab terminations. Chemically complex or electronically demanding minerals, such as Fe-containing phases, therefore benefit particularly from MLIPs when many candidate surfaces must be relaxed and ranked.

Scaling tests on an NVIDIA GeForce RTX 4090 GPU show that both the qNEP version and the conventional NEP version retain high computational throughput as system size increases (**Fig. 3g**). Enabled by the particle–particle particle–mesh (PPPM) method, qNEP retains high computational efficiency relative to conventional NEP while incorporating environment-dependent charge effects[31]. This scalability enables high-throughput screening of candidate facets and terminations, as well as the construction of large, application-ready surface models for molecular dynamics simulations across energy, environmental and advanced inorganic interfaces.

**Wulff-based validation and prediction of mineral surface exposure**

Wulff construction provides the thermodynamic link between surface-energy landscapes and equilibrium crystal morphology, allowing stable facets to be identified according to their expected surface exposure[33,34]. Accurate surface-energy prediction becomes useful for surface model selection only when the predicted energies are translated into the facets that are expected to remain exposed. We therefore used the MinNEP-predicted surface energies to construct equilibrium Wulff morphologies and quantify surface area fractions. This morphology-level analysis provides a more stringent assessment than individual surface-energy errors alone, because Wulff construction depends on the relative energetic hierarchy among facets. Small changes in this hierarchy can alter whether a surface persists, shrinks or disappears from the equilibrium morphology.

α-Quartz was used as a representative benchmark to evaluate whether MinNEP preserves the morphology-determining surface-energy hierarchy. The DFT- and MinNEP-derived Wulff constructions show similar equilibrium shapes and retain the same dominant exposed surface families (**Fig. 4a,b**). Although the exact surface area fractions differ slightly, MinNEP preserves the facets that control the equilibrium morphology. This agreement indicates that the model reproduces not only the magnitude of individual surface energies, but also the relative stability relationships required for Wulff-based surface exposure prediction.

We further compared the MinNEP-resolved quartz surfaces with experimental XRD measurements. The measured diffraction pattern exhibits characteristic peaks associated with the predicted crystallographic planes, providing an experimental crystallographic check for the resolved surface model (**Fig. 4c**). Because powder XRD reflects long-range crystallographic order rather than directly measuring surface area fractions, this comparison should be interpreted as supporting consistency between the predicted exposed planes and the experimentally observed quartz structure, rather than as a direct quantification of equilibrium morphology. Together with the DFT Wulff comparison, it supports the use of MinNEP-derived Wulff morphologies as physically grounded references for constructing quartz surface models.

The workflow was then applied to the remaining representative minerals considered in this study. The resulting Wulff constructions reveal distinct exposure patterns across dicalcium silicate ($C_2S$), tricalcium silicate ($C_3S$), tricalcium aluminate ($C_3A$), tetracalcium aluminoferrite ($C_4AF$), wollastonite, albite, anorthite, orthoclase and microcline (**Fig. 4d**). In several cases, only a small subset of the enumerated slab terminations contributes appreciably to the final morphology, whereas higher-energy surfaces vanish or remain weakly expressed. This result shows that stable surface exposure cannot be inferred reliably from Miller indices alone or from a small number of conventional cleavage planes. Instead, morphology-relevant surfaces emerge from the surface-energy hierarchy across the enumerated termination space.

These Wulff morphologies provide the final link between surface-energy prediction and downstream atomistic modelling[35–37]. Rather than selecting mineral slabs empirically, MinSurf identifies the facets and terminations that are thermodynamically favourable and morphologically exposed for each mineral. The resulting surface models can serve as standardized inputs for molecular dynamics simulations of hydration, dissolution, adsorption and interfacial bonding. Because additional minerals, facets and surface chemistries can be incorporated into the same workflow, MinSurf provides an expandable route for constructing physically grounded mineral surface models beyond the specific systems examined here.

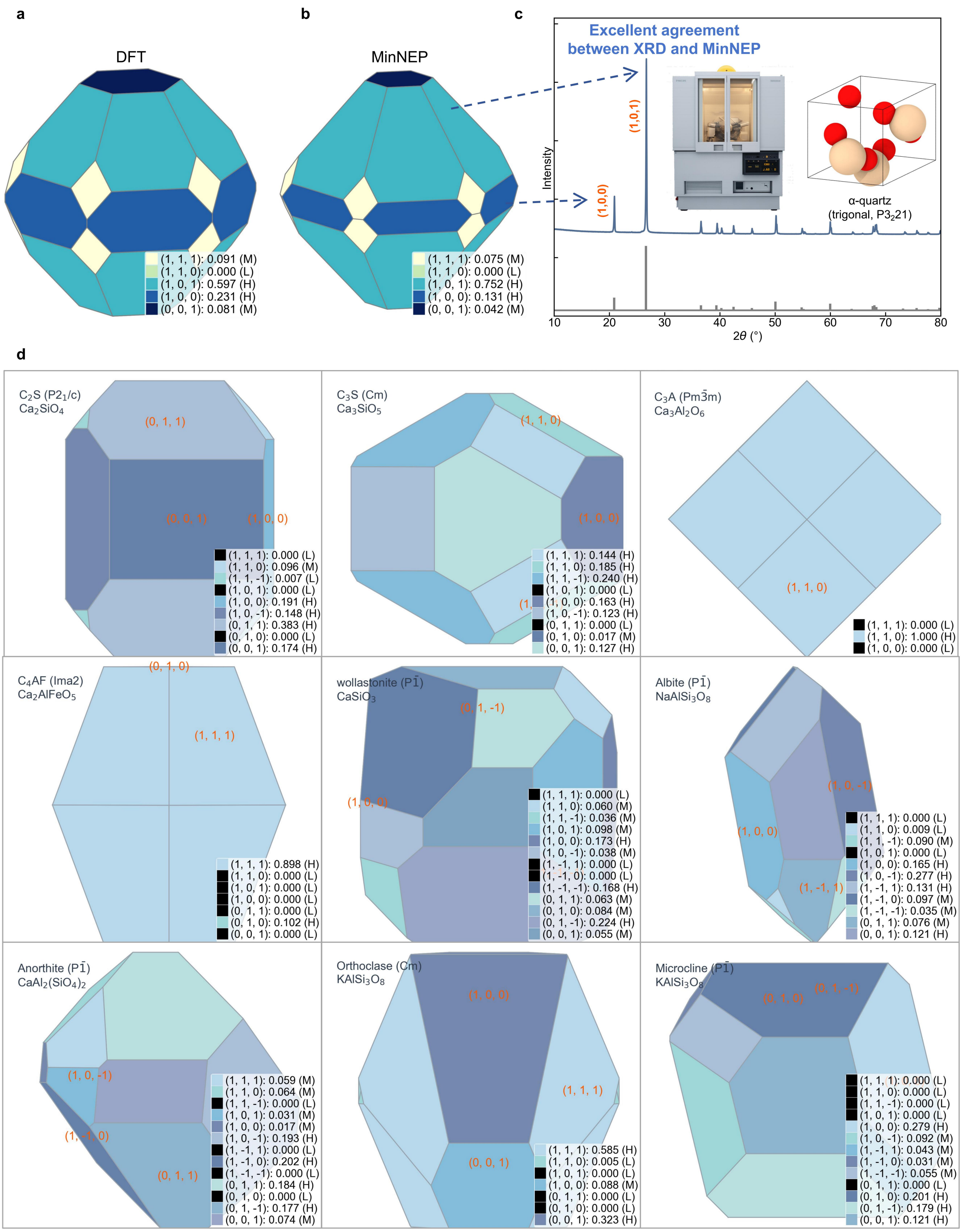

**Fig. 4 | Wulff-based validation and application of MinNEP-predicted mineral surface exposure. a,b,** Comparison of DFT- and MinNEP-derived Wulff constructions for α-quartz, with surface area fractions and relative surface-energy levels labelled for each exposed facet. **c,** Experimental XRD comparison of α-quartz, showing consistency between characteristic diffraction features and the predicted crystallographic planes. **d,**

MinNEP-predicted Wulff constructions for representative minerals, with exposed Miller indices and surface area fractions summarized in each panel.

## Discussion

This work reframes mineral surface construction as a problem of resolving the surface landscape of a crystal, rather than as a matter of selecting a conventional slab. In this view, surface identity is not determined by a Miller index alone, but emerges from the combined effects of crystallographic orientation, atomic termination, stoichiometry, polarity, relaxation and morphological exposure[38]. Distinct atomic surfaces associated with the same orientation can occupy different positions in the surface-energy landscape and can therefore contribute differently to the equilibrium morphology[39]. By enumerating these candidates and evaluating them through a common energetic and morphological criterion, MinSurf provides a reproducible route for translating crystallographic surface possibilities into morphology-aware atomic surface models.

This shift has direct implications for molecular simulations of mineral interfaces. The relevant input for downstream simulations is not simply a plane such as (001) or (010)[14], but a relaxed atomic configuration with a defined bulk reference, energetic stability and expected exposure. If a slab is not representative of a stable or morphologically exposed surface, simulated interfacial processes may reflect the modelling choice as much as the intrinsic chemistry of the mineral. MinSurf addresses this problem by linking atomistic slab construction, surface-energy landscapes and Wulff-derived exposure[40]. The α-quartz comparison illustrates why morphology-level agreement is a stricter test than isolated surface-energy accuracy: a model must preserve the relative energetic hierarchy that determines which facets remain exposed, rather than only reproduce individual surface-energy values.

Resolving this landscape at practical scale requires models that are both efficient and physically expressive. Mineral surfaces commonly involve under-coordinated atoms, distorted coordination polyhedra, polar cuts and partially ionic bonding, making surface relaxation and termination stability sensitive to electrostatic interactions and local charge redistribution[38]. Purely local models can capture many short-range structural correlations, but may require larger cutoffs or more extensive training data when long-range electrostatic contributions affect the relative stability of competing surfaces. MLIPs that combine local structural descriptors with environment-dependent charges or explicit long-range electrostatic terms, such as the formulation used in MinNEP, provide a more suitable representation for these partially ionic surface environments. Their key value is not merely lower prediction error, but the ability to make landscape-level screening feasible by relaxing, ranking and converting large numbers of candidate surfaces into morphology-aware models.

MinSurfSet is more than a training set for the present model. It provides a structured representation of mineral surface landscapes, in which each slab is linked to a parent crystal, an OUC reference, a Miller index, an atomic termination and a DFT label. This structure makes MinSurfSet suitable for systematic expansion as new minerals, higher-index facets, surface chemistries or chemical environments are introduced. More importantly, it positions MinSurfSet as a morphology-aware benchmark dataset for future MLIPs for mineral surfaces. Such a benchmark should test not only energy and force accuracy, but also whether a model preserves surface-energy hierarchies, termination stability and Wulff-derived exposure, because small errors in relative surface energies can change the predicted exposed morphology. This makes MinSurfSet more than a repository of slab energies and forces: it provides a benchmark for whether an MLIP can support physically meaningful surface selection.

Several limitations define the next stage of MinSurf. The present workflow focuses mainly on dry surfaces generated from crystalline bulk structures, whereas real mineral interfaces may be hydroxylated, hydrated, defective, reconstructed or covered by ions and organic molecules[41,42]. These effects can modify both absolute surface energies and relative termination stability. Wulff construction provides an equilibrium thermodynamic reference, but does not capture kinetic effects associated with fracture, grinding, growth, dissolution or ageing[40]. The XRD comparison for α-quartz provides crystallographic consistency, not a direct measurement of surface area fractions. Extending MinSurfSet to hydrated, hydroxylated, defective and adsorbate-covered surfaces, together with MLIPs that explicitly incorporate long-range interactions or physics-informed constraints, would allow mineral surface landscapes to be resolved under more realistic chemical environments. Overall, by translating first-principles surface landscapes into reproducible atomic models, MinSurf provides a standardized and expandable basis for more reliable simulations of mineral interfaces across natural, engineered and planetary materials.

## Methods

### Crystal structure preparation and surface model generation

The initial crystal structures were collected from crystallographic databases and materials repositories. Ten mineral crystals were selected in this work, including dicalcium silicate ($Ca_2SiO_4$), tricalcium silicate ($Ca_3SiO_5$), tricalcium aluminate ($Ca_3Al_2O_6$), tetracalcium aluminoferrite ($Ca_2AlFeO_5$), wollastonite ($CaSiO_3$), α-quartz ($SiO_2$), albite ($NaAlSi_3O_8$), anorthite ($CaAl_2(SiO_4)_2$), orthoclase ($KAlSi_3O_8$) and microcline ($KAlSi_3O_8$). $C_2S$ and $C_4AF$ were obtained from the American Mineralogist Crystal Structure Database[26], $C_3S$ was obtained from the Cambridge Crystallographic Data Centre[24], and the remaining structures were obtained from the Materials Project[25].

OUCs and slab models were constructed from the collected crystal structures using a surface-orientation-based slab generation method implemented in the pymatgen package[30,43,44]. Candidate facets were enumerated from Miller indices satisfying |h|, |k| and |l| ≤ 1, excluding the (0, 0, 0) plane. For each selected orientation, an OUC was constructed as the bulk reference, and candidate surface slabs were generated by sampling different termination shifts. Each surface model was identified by its mineral identity, Miller index and termination shift. The slab thickness was set to no less than 10 Å, and a vacuum layer of at least 15 Å was introduced along the surface-normal direction. This procedure generated 90 OUCs and 764 candidate surface slabs in total.

**Density functional theory calculations**

DFT calculations were performed using the Vienna Ab initio Simulation Package. The projector augmented-wave method and the Perdew–Burke–Ernzerhof exchange–correlation functional within the generalized gradient approximation were used[45,46]. The plane-wave energy cutoff was set to 520 eV. The precision level was set to Normal, and real-space projection was evaluated automatically. Electronic self-consistency was converged to $10^{-6}$ eV. Gaussian smearing was used with a smearing width of 0.1 eV. Spin-polarized calculations were performed for Fe-containing systems, whereas spin-unpolarized calculations were used for all other systems.

Geometry optimizations were carried out for mineral crystals, OUCs and surface slabs using structure-specific settings. Mineral crystals were relaxed with variable cell shape and volume until the residual force was below 0.01 eV $Å^{-1}$. OUCs and surface slabs were relaxed with fixed cell parameters and a force convergence criterion of 0.02 eV $Å^{-1}$. The quasi-Newton algorithm was used for mineral crystals and OUCs, with the conjugate-gradient algorithm used when required for robust convergence. Surface slabs were relaxed using the conjugate-gradient algorithm.

The energies and atomic forces generated during structural optimizations were collected as DFT reference labels for MinSurfSet. Γ-centred k-point meshes were used throughout. The k-point density was tested from 5 to 35, and the selected density was defined as the lowest value for which all higher-density calculations remained within 1 meV $atom^{-1}$ of the highest-density reference calculation. For surface slabs, the same in-plane k-point sampling was used, whereas the surface-normal direction was sampled only at Γ. Detailed input parameters and k-point convergence tests are provided in the **Supplementary Sections S3** and **S4**.

**Machine-learning interatomic potential architecture**

We developed a charge-aware neuroevolution potential, termed MinNEP, using the qNEP framework implemented in GPUMD v4.8[31,47,48]. The qNEP architecture was adopted because it retains the computational efficiency of conventional NEP while introducing environment-dependent effective charges and long-range electrostatic interactions. In MinNEP, the total energy is represented by a local NEP contribution, a reciprocal-space electrostatic contribution derived from the learned charges and a short-range repulsive correction[49]. This formulation is suitable for mineral surface systems, where cleavage and relaxation can generate under-coordinated atoms, polar environments and local charge redistribution. Detailed model principles are provided in **Supplementary Section S1**.

MinNEP was trained for seven elements, Ca, Si, O, Na, Al, Fe and K, using the fourth-generation NEP formalism. Both radial and angular cutoff radii were set to 7 Å. Charge mode 2 was used, in which only the reciprocal-space electrostatic contribution is explicitly retained, reducing possible double counting of short-range electrostatic interactions already represented by the local NEP term. A Ziegler–Biersack–Littmark correction with a cutoff of 2.5 Å was included to describe short-range repulsion. The model was optimized against DFT energies and atomic forces with equal weights, whereas virial loss and L1 regularization were disabled. The batch size was set to 200, and training was performed for 200,000 generations. Training details and hyperparameter settings are provided in **Supplementary Sections S5 and S6**.

**Surface energy calculation**

Surface energies were evaluated using the generated paired OUCs and slab models. All MinNEP-based calculations were performed using GPUMD 4.8[47,48]. For each surface indexed by (*hkl*), the OUC was used as the bulk reference, and the corresponding slab was used to represent the exposed surface. During slab relaxation, atoms in the middle region of the slab, defined as those located more than 2 Å away from both exposed surfaces, were fixed to preserve a bulk-like interior, whereas surface atoms were allowed to relax.

Structural optimizations were performed using the steepest descent optimizer until the force convergence threshold of $10^{-6}$ eV Å$^{-1}$ was reached, or until no further reduction in system energy could be achieved. The average surface energy of each (*hkl*) slab was calculated as:

$$E_{surf}^{hkl,\tau} = \frac{E_{slab}^{hkl,\tau} - N^{hkl,\tau} E_{\mathrm{OUC}}^{hkl,\tau}}{2A^{hkl,\tau}}$$

where $E_{surf}^{hkl,\tau}$ is the surface energy, $E_{slab}^{hkl,\tau}$ is the total energy of the relaxed slab, $E_{\mathrm{OUC}}^{hkl,\tau}$ is the bulk OUC energy per atom, $N^{hkl,\tau}$ is the number of atoms in the slab, and $A^{hkl,\tau}$ is the surface cross-sectional area, all corresponding to termination $\tau$ of the (*hkl*).

**Wulff construction**

The equilibrium morphology of the mineral crystal was predicted using the Wulff construction based on the calculated orientation-dependent surface energies. Rather than treating the exposed surfaces independently, this method determines the stable crystal shape by imposing a geometric relationship between the surface energy of each crystallographic plane and its distance from the crystal center. For a given (*hkl*) surface, the Wulff condition can be expressed as:

$$\frac{d_{hkl}}{\gamma_{hkl}} = \lambda$$

where $\gamma_{hkl}$ is the surface energy of the (*hkl*) facet, $d_{hkl}$ denotes the perpendicular distance from the crystal center to this facet, and $\lambda$ is a proportionality constant determined by the fixed crystal volume. According to this relationship, facets with lower surface energies tend to be more prominently expressed in the equilibrium morphology, whereas high-energy facets are less stable and may contribute only a small surface area or disappear from the final Wulff shape. The resulting Wulff morphology was then used to quantify the relative area fraction of each exposed crystallographic surface of the mineral crystal.

**Experimental XRD characterization**

Natural quartz from Gongyi, Henan Province, China, was used for experimental validation. The sample had a $SiO_2$ content of approximately 99.56 wt%, as determined by inductively coupled plasma optical emission spectroscopy. The raw quartz was pulverized using a planetary ball mill to a particle size below 5 μm and sieved through a 3500-mesh standard screen before measurement. Powder X-ray diffraction was performed on an X'Pert PRO diffractometer (PANalytical, Netherlands) using Cu Kα1 radiation (λ = 1.5406 Å) at 40 kV and 40 mA. Diffraction patterns were collected in continuous scan mode over a 2θ range of 5.0–90.0°, with a counting time of 97.92 s per data point.

## Data availability

All of the data supporting this study are available at repository via https://huggingface.co/datasets/Fengzijun/MinSurfProjectV1.0.

## Code availability

All codes developed to support this study are available from the repository at https://huggingface.co/datasets/Fengzijun/MinSurfProjectV1.0.

MinNEP training and calculations were performed using the open-source GPUMD package[47], which is developed and maintained independently and is available at https://github.com/brucefan1983/GPUMD.

## Acknowledgements

The authors are grateful for financial support from the National Natural Science Foundation of China (52308448).

## Competing interests

The authors declare no competing interests.

## Additional information

*SupplementaryInformation.pdf* is available for this paper.